\documentclass[letter,oldversion]{aa} 
\usepackage{graphicx}
\usepackage{txfonts}
\usepackage{rotating}
\usepackage{lscape}
\usepackage{natbib}
\newcommand{\gsim}{\;\lower.6ex\hbox{$\sim$}\kern-7.75pt\raise.65ex\hbox{$>$}\;}
\newcommand{\lsim}{\;\lower.6ex\hbox{$\sim$}\kern-7.75pt\raise.65ex\hbox{$<$}\;}

\begin{document}
\title{NGC~6388 reloaded: some like it hot, but not too much\thanks{Based on
observations collected at  ESO telescopes under programmes 073.D-0211,
073.D-0760, 381.D-0329, 095.D-0834, and 099.D-0047}
 }

\author{
Eugenio Carretta\inst{1}
\and
Angela Bragaglia\inst{1}
}

\authorrunning{Carretta and Bragaglia}
\titlerunning{Abundances in NGC~6388}

\offprints{E. Carretta, eugenio.carretta@inaf.it}

\institute{
INAF-Osservatorio di Astrofisica e Scienza dello Spazio di Bologna, via Gobetti
93/3, I-40129 Bologna, Italy}

\date{}

\abstract{Multiple stellar populations in globular clusters (GCs) are defined
and recognized by their chemical signature, with second generation stars showing
the effects of nucleosynthesis in the more massive stars of the earliest
component formed in the first star formation burst. High temperature H-burning
produces the whole pattern of (anti)-correlations among proton-capture elements
widely found in GCs. However, where this burning occurred is still debated. Here
we introduce new powerful diagnostic plots to detect evidence (if any) of
products from proton-capture reactions occurring at very high temperatures.
To test these Detectors Of High Temperature (in short DOHT) H-burning plots 
we show how to put stringent constraints on the temperature range of
the first generation polluters that contributed to shape the chemistry of
multiple stellar population in the massive bulge GC \object{NGC~6388}. Using the largest
sample to date (185 stars)  of giants with detailed abundance ratios in a single
GC (except $\omega$ Cen) we may infer that the central temperature of part of the
polluters must have been comprised between $\sim$100 and $\sim$150 million Kelvin (MK) if we
consider hydrostatic H-burning in the core of massive stars. A much narrower
range (110 to 120 MK) is inferred if the polluters can be identified in massive
asymptotic giant branch (AGB) stars.
}

\maketitle

\section{Introduction}
The multiple stellar populations in Galactic globular clusters (GCs) are
characterized by their pattern of light elements (C, N, O, Na, Mg, Al, Si, and
sometime heavier species such as Ca, Sc, and K). One component, which is believed to be the first to be
formed in a proto-GC, presents the same level of abudance ratios as the vast
majority of metal-poor halo stars at the same metallicity. However, the so
called second-generation (SG) shows alterations in the abundances of these
elements that range from moderate to extreme. The network of
correlations and anti-correlations found among these species represents a key
clue to explain the origin of multiple stellar population, since all the
observed relations can be traced back to one single mechanism, H-burning.

Since the pioneering studies by \citet{dd89}  and \citet{langer93} it was clear that proton-capture reactions in H-burning at high
temperature could account for all the chemical modifications observed. The
enhancement in the abundance ratios of a given element is accompanied (and
explained) by the depletion of some other species participating to the network
of proton-capture reactions.
The uncertainties related to the modeling of stellar structures and to the
release of processed material into the intracluster medium hamper at present
the knowledge of the exact sites where this burning occurred. Several candidate
polluters were proposed, ranging from the less massive ones (intermediate-mass
asymptotic giant branch stars, AGB; \citealt{CDC81,ventura01}) up to supermassive objects \citep{DH14}, passing from
fast-rotating massive stars \citep{decressin07} or interacting massive
binaries \citep{demink09}.

Fortunately, the variety of nuclear reactions involved in this burning provides
some clues, since proton-capture on heavier and heavier nuclei must account for
the increasing Coulomb barrier. As a consequence, higher and higher temperatures
are required to activate the relevant reactions, which thus provide ideal
thermometers to probe the inner structure of the stellar sites where the
nucleosynthesis occurred. 
Thus, fusion temperatures $\geq 40$ MK are enough for the conversion of O into N
and the production of Na from the NeNa cycle to work simultaneously. This
rather modest threshold  explains why the Na-O anticorrelation is so widespread
among GCs that is was proposed to be considered the main signature of the
presence of multiple populations, in other words the genuine essence of a
globular cluster \citep{carretta10}. The higher temperatures ($\geq 70$ MK)
necessary for the MgAl cycle to efficiently operate, apparently are possible only
among stars in a subset of GCs, because the Mg-Al anti-correlation was only found
in massive and/or metal-poor GCs \citep{carretta09a}, as later confirmed by
further studies \citep{meszaros15,pancino17,nataf19} . These are also the GCs where some overproduction of Si is observed \citep[e.g.][]{carretta09a}, pointing out that temperatures in excess of 80-100 MK were
attained in the polluters \citep[e.g.][]{prantzos17}. 
Finally, even higher temperatures $>100-180$ MK \citep{ventura12,prantzos17} are required to explain the Mg-K and Mg-Sc anticorrelations,
including some of the heaviest proton-capture species and observed 
in a couple of peculiar GCs: NGC~2808 \citep{mucciarelli15,carretta15} 
and NGC~2419 \citep{mucciarelli12,cohen12}.

In the present Letter we exploit these different thermometers to design new
diagnostic plots that we named Detectors Of High Temperature (DOHT) H-burning
products, since they allow to reveal the effects (if any) of reactions activated
at unusually high temperatures, exceeding the ones tipically occurring in the 
majority of clusters' polluters. We test the DOHT plots by putting strong
constraints on the temperature range of the putative polluters that shaped the
chemical pattern of multiple populations in the massive GC NGC~6388.

Data and analysis are briefly summarized in Section 2, whereas results are
presented in Section 3 and discussed in Section 4.

\section{The dataset}

NGC~6388 is a bulge cluster of high total mass, yet only 37 stars with
determined abundances of proton-capture elements were available 
\citep{carretta07,carretta09b}, due to the high field contamination affecting this GC. This
problem was recently overcome \citep[][hereinafter Paper I]{CB18}
 by exploiting the large sample of stellar spectra in the
cluster region present in the ESO archive as well as newly granted 
observing time. Here we present results based on homogeneous abundances of Mg,
Ca, Sc of more than 180 giants, the largest sample of stars with detailed
abundance analysis in a GC (with the notable exception of $\omega$ Cen). This
dataset was used to test the new DOHT plots, as shown in the next Section.

Our results are based on abundances partly from UVES spectra presented
in Paper I (23 stars), partly on new UVES spectra (11 stars), and finally on 149
stars with abundances of Fe, Mg, Ca, Sc from GIRAFFE spectra and the
high-resolution setup HR13. We acquired new
data with the ESO programme 099.D-0047 (PI Carretta) and used archival spectra 
of three programmes: 381.D-0329(B), PI Lanzoni;  073.D-0760(A), PI
Catelan; 095.D-0834(A), PI Henault-Brunet, plus the original data from our
FLAMES survey (073.D-0211, PI Carretta). The detailed abundance analysis for
all stars not presented in Paper I will be described in a forthcoming paper
(Carretta and Bragaglia 2019a). However, it closely follows our
homogeneous procedures (see \citealt{carretta09a,carretta09b} for the UVES and GIRAFFE
spectra, respectively) for deriving abundances and estimating star-to-star
errors. The metallicity for NGC~6388 is [Fe/H]$=-0.480$ dex (rms=0.045 dex,
35 stars), based on UVES spectra \citep{CB19a} and on the same method used to
derive the metal abundance for NGC~2808 ([Fe/H]$=-1.129$ dex, rms=0.030 dex, 31
stars, \citealt{carretta15}), that we will use as a comparison in the next
Section.

\section{Temperature range of FG polluters in NGC~6388}

\begin{figure}
\centering
\includegraphics[scale=0.40]{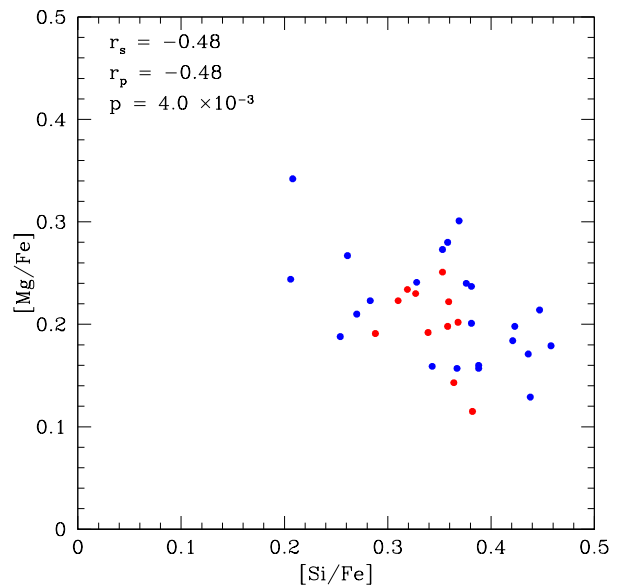}
\caption{Anti-correlation of [Mg/Fe] and [Si/Fe] abundance ratios in NGC~6388.
Blue and red circles indicate stars from Paper I and the new stars analyzed in
Carretta and Bragaglia (2019a), respectively. The Spearman rank correlation
($r_S$) coefficient, the Pearson linear correlation ($r_P$) coefficient and its
p-value are listed.}
\label{f:summgalsi}
\end{figure}

The first constraint simply expands over the result already shown in Paper I,
and based on abundances from UVES spectra. In Figure~\ref{f:summgalsi}
we show the Mg-Si anti-correlation in NGC~6388, which is found to be
statistically very significant: the Pearson regression coefficient $r_P$ is
-0.48 (34 stars), with a two-tail probability to be a random result $p = 4.0
\times 10^{-3}$.

This is clearly due to the well known leakage from Mg-Al cycle on $^{28}$Si: 
when the $^{27}$Al($p$,$\gamma$)$^{28}$Si reaction takes over
$^{27}$Al($p$,$\alpha$)$^{24}$Mg a certain amount of $^{28}$Si is produced
by proton-captures \citep[see][]{KL03}. This overproduction of Si occurs typically at temperatures
$> 100$ MK, even if the leakage from the Mg-Al cycle already starts at about 65
MK \citep{arnould99}. The first constraint then stems from these
observations, strenghtened and well assessed by the analysis of 11 new red
giants with UVES spectra in Carretta and Bragaglia (2019a; red points in 
Figure~\ref{f:summgalsi}). 

\begin{figure}
\centering
\includegraphics[scale=0.40]{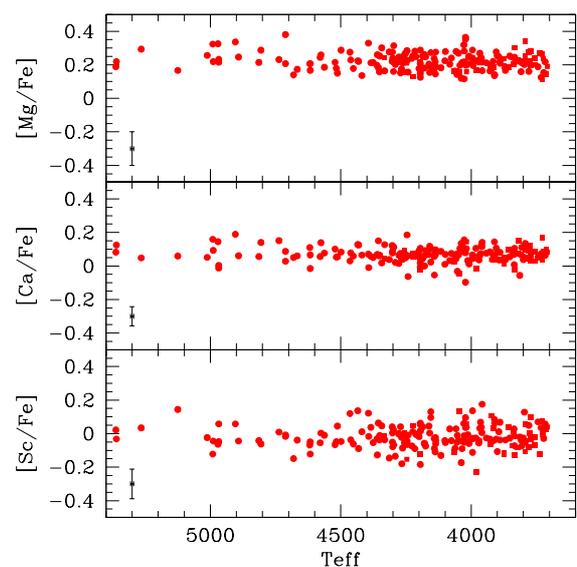}
\caption{The ratios [Mg/Fe] (184 stars), [Ca/Fe], and [Sc/Fe] (185 stars)  from
FLAMES spectra in NGC~6388 (Carretta and Bragaglia 2019a) as a function of
effective temperature. Internal error bars are shown in each panel.}
\label{f:teff3}
\end{figure}

A second limit to the temperature range in FG polluters active at the cluster
formation may be provided by looking at the abundances of Mg, Ca, and Sc,
available for 184, 185, and 185 stars, respectively, in NGC~6388 (Carretta and
Bragaglia 2019a). The abundance ratios are plotted in Figure~\ref{f:teff3} as a
function of the effective temperatures. Star to star error bars are also shown
and estimated (as described in \citealt{carretta09b} as 0.100 dex, 0.057 dex,
and 0.088 dex for [Mg/Fe], [Ca/Fe], and [Sc/Fe], respectively.

\begin{figure*}
\centering
\includegraphics[scale=0.40]{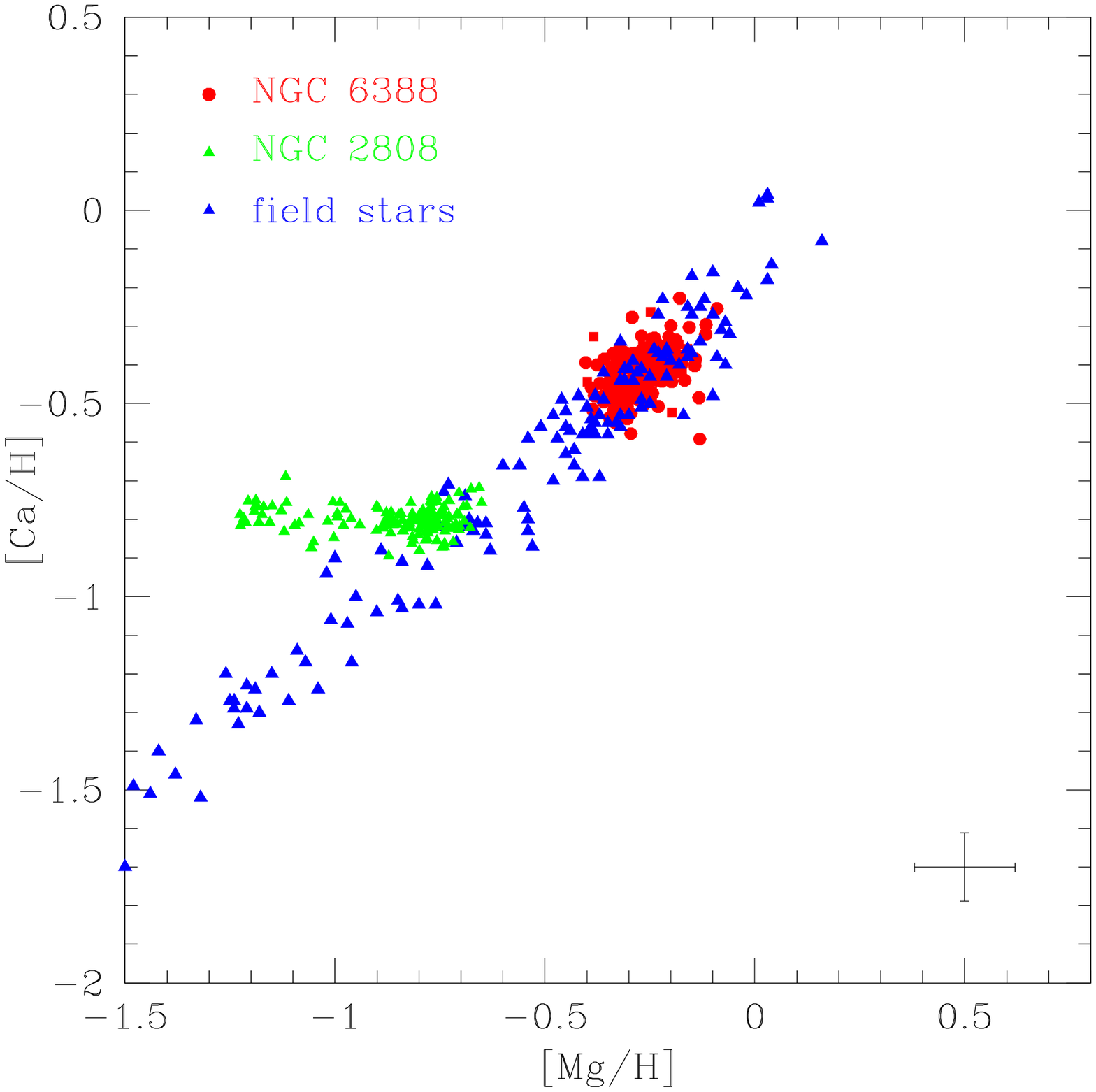}\includegraphics[scale=0.40]{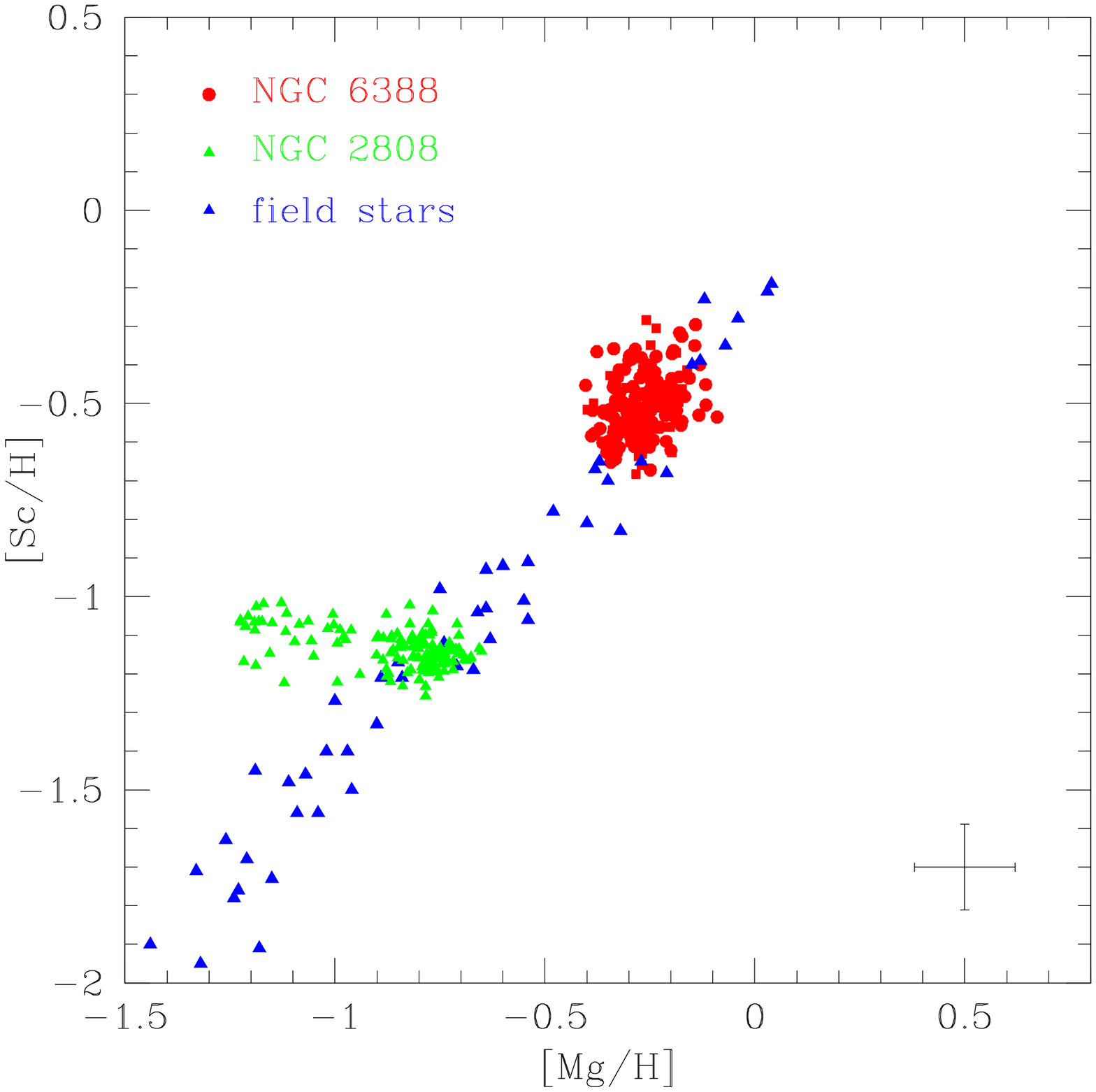}
\caption{The ratios [Ca/H] (left panel) and [Sc/H] (right panel) as a function
of [Mg/H] for giants in NGC~6388 (red points), superimposed to field stars from
Gratton et al. (2003; blue triangles). Green triangles are giants in NGC~2808
from Carretta (2015). Internal star-to-star errors refer to NGC~6388.}
\label{f:cahsch}
\end{figure*}

The constraint on the H-burning temperature in the stellar sites that forged the
proton-capture elements ending up in forming the multiple populations in
NGC~6388 may be fully appreciated from the two panels in 
Figure~\ref{f:cahsch}, that we name DOHT plots. In this Figure we used field
stars from \citet{gratton03} as reference, since the majority
of the field stars only show the chemical pattern derived from the Supernovae
(SNe) nucleosynthesis, with no trace of the peculiar proton-capture
nucleosynthesis which is restricted to the dense environment of GCs \citep[e.g.][]{gratton00,SM03}.

The distribution of stars of NGC~6388 (red points) in the Ca-Mg and Sc-Mg planes
matches almost perfectly that of field stars: Ca and Sc abundances in this GC
are not modified by the action of proton-capture reactions. 

Conversely, the observations present a different scenario in NGC~2808, a GC
slighly less massive than NGC~6388 (total absolute magnitude $M_V=-9.39$ and
$M_V=-9.41$ mag, respectively: \citealt{harris96}, on-line version 2010). The same
abundances are plotted (green triangles) in the two panels of 
Figure~\ref{f:cahsch} and all elemental ratios are obtained with extremely
homogeneous procedures, the same adopted also for NGC~6388. The pattern observable in
NGC~2808 is strikingly different, with abundances of both Ca and Sc being
anti-correlated to abundances of Mg. Both these relations are found to be
statistically very significant \citep{carretta15}. This behaviour was attributed to
the same process of proton-capture at very high temperature (above 100 MK)
postulated to act in GCs like NGC~2419 to produce potassium from proton capture
on $^{36}$Ar nuclei \citep{ventura12}. The observations in NGC~2808 are
explained with this high temperature process, where the formation of proton-rich
species is shifted toward heavier nuclei. The effect are obviously more marked
on Sc (and K) than Ca, the latter being much more abundant.

The inference for the case of NGC~6388 is straightforward: the H-burning in the
FG polluters in this GC was not able to reach temperatures high enough to
substantially modify the abundances of Ca and Sc with respect to the levels
from SNe nucleosynthesis. 

\section{Discussion and conclusions}

Summarizing the observational evidence found in NGC~6388, we notice (i)
overproduction of Si likely due to the increased leakage from the Mg-Al cycle
(Paper I and the present letter, Figure~\ref{f:summgalsi}), and (ii) absence of
significant Ca-Mg and Sc-Mg anticorrelations in the new diagnostic plot 
introduced in the present work (Figure~\ref{f:cahsch}).

\citet{prantzos17} made extensive H-burning nucleosynthesis calculations
in the range of temperature and density suitable to hydrostatic burning in the
H-shell of AGB stars or in the cores of massive and supermassive stars.
After accounting for the minimum dilution factor required to reproduce the most
extreme O abundances in NGC~2808, they found that a limited temperature range
(70-80 MK) is able to reproduce all the extreme observed values of O, Na, Mg,
and Al in this cluster. The high estreme of this range also accounts for the
observed Si abundances in NGC~2808. 
Only the burning temperature required for K production exceeds this range, since
the extreme K abundances in NGC~2808 only are reproduced for T$\sim 180$ MK. A
similar discussion can be found in \citet{dantona16}.

In NGC~6388 we do not have (yet) determinations of [K/Fe], but we can address
the production of the heavy proton-capture elements by looking at Ca and Sc,
with similar atomic numbers, whose abundances are also clearly  altered by
H-burning in FG polluters of NGC~2808 \citep{carretta15}. At odds with the latter
case, in the present study we found from the DOHT plots that in NGC~6388 the distribution of Ca and
Sc closely follows that of field stars at similar metallicity. Conversely, we
see robust evidence that Si is altered (Paper I for the UVES sample and Carretta
and Bragaglia 2019a for the GIRAFFE sample), implying temperatures
higher than 100 MK. Since heavier elements (like Ar and K) do not start to be
affected by nuclear burning before T $\sim 150$ MK \citep{prantzos17}, our
present results allow to pinpoint a rather limited range of temperature ($\sim
100 - 150$ MK) where the H-burning in the putative FG polluters occurred. An
even narrower range can be estimated if the polluters can be identified in
massive AGB stars. In Figure 1 of \citet{dantona16}, at metallicities
corresponding to that of NGC~6388, the maximum temperature T$_{HBB}$ at the
bottom of the convective envelope for masses evolving in AGB exceeds the one
necessary to produce Si, but never reaches the one required to process Ar into K
or heavier species. In this case a range 110-120 MK may explain the observed
abundances in NGC~6388.

However, already with the limited sample of stars available in Paper I there was
some hints that a single source of pollution is probably not enough to fit all
the multi-elements observations with a dilution model in NGC~6388.  
More stringent conclusions will be drawn when the full set of light
elements (O, Na, Mg, Al, Si, with Ca and Sc as the heaviest extreme) will be
available. 

The present Letter highlights an application to infer the temperature
range at which past polluters were active and illustrates how much precise could
be the constraints potentially available from large samples with a complete set 
of homogeneous abundances of proton-capture elements from spectroscopy.
The new diagnostic DOHT plots seem to be a powerful tool to test the existence of
products from nuclear processing at very high temperatures. The GCs where this
very advanced H-burning operated are currently rare, but their census is
growing. Among these GCs we may already annoverate NGC~2808 \citep{carretta15},
NGC~2419 \citep{mucciarelli12, cohen12}, and NGC~4833 (Carretta
and Bragaglia 2019b).

Datasets like the one used here and in the forthcoming paper can be used by
theoreticians to test and tune both models and derived scenarios up to
moderately high metal abundances. A set of calculations for NGC~6388, like those
presented by \citet{dantona16} and \citet{prantzos17} for NGC~2808 and
 \citet{prantzos07} for NGC~6752, would be very welcome, especially because
the abundances in all these three GCs, spanning 1 dex in [Fe/H], are obtained in
a very homogeneous way.

\begin{acknowledgements}
We thank Raffaele Gratton for useful discussions. This research has made use of the services of the ESO Science Archive Facility,
of the SIMBAD database (in particular  Vizier), operated at CDS, Strasbourg,
France, and of the NASA's Astrophysical Data System.
\end{acknowledgements}

\end{document}